# Van-Hove annihilation and nematic instability on a Kagome lattice

**Authors:** Yu-Xiao Jiang[1]†*, Sen Shao[3]*, Wei Xia[4,8]*, M. Michael Denner[2]*, Julian Ingham[9]*, Md Shafayat Hossain[1], Qingzheng Qiu[10], Xiquan Zheng[10], Hongyu Chen[3], Zi-Jia Cheng[1], Xian P. Yang[1], Byunghoon Kim[1], Jia-Xin Yin[5], Songbo Zhang[2], Maksim Litskevich[1], Qi Zhang[1], Tyler A. Cochran[1], Yingying Peng[10], Guoqing Chang[3], Yanfeng Guo[4,8], Ronny Thomale[6], Titus Neupert[2], M. Zahid Hasan[1,7]†

**Affiliations:**

[1]Laboratory for Topological Quantum Matter and Advanced Spectroscopy, Department of Physics, Princeton University, Princeton, New Jersey, USA.

[2]Department of Physics, University of Zürich, Winterthurerstrasse 190, 8057, Zürich, Switzerland

[3]Division of Physics and Applied Physics, School of Physical and Mathematical Sciences, Nanyang Technological University, Singapore 639798, Singapore.

[4]School of Physical Science and Technology, ShanghaiTech University, Shanghai 201210, China

[5]Department of physics, Southern University of Science and Technology, Shenzhen, Guangdong 518055, China.

[6]Institut für Theoretische Physik und Astrophysik, Julius-Maximilians-Universität Würzburg, Würzburg, Germany.

[7]Quantum Science Center, Oak Ridge, Tennessee 37830, USA.

[8]ShanghaiTech Laboratory for Topological Physics, ShanghaiTech University, Shanghai 201210, China

[9]Department of Physics, Columbia University, New York, NY, 10027, USA

[10]International Center for Quantum Materials, School of Physics, Peking University, Beijing 100871, China

**Abstract:**

Novel states of matter arise in quantum materials due to strong interactions among electrons. A nematic phase breaks the point group symmetry of the crystal lattice and is known to emerge in correlated materials. Here we report the observation of an intra-unit-cell nematic order and signatures of Pomeranchuk instability in the Kagome metal ScV6Sn6. Using scanning tunneling microscopy and spectroscopy, we reveal a stripe-like nematic order breaking the crystal rotational symmetry within the Kagome lattice itself. Moreover, we identify a set of van Hove singularities adhering to the Kagome layer electrons, which appear along one direction of the Brillouin zone while being annihilated along other high-symmetry directions, revealing a rotational symmetry breaking. Via detailed spectroscopic maps, we further observe an elliptical deformation of Fermi surface, which provides direct evidence for an electronically mediated nematic order. Our work not only bridges the gap between electronic nematicity and Kagome physics, but also sheds light on the potential mechanism for realizing symmetry-broken phases in correlated electron systems.

**Main text:**

Spontaneous symmetry breaking is a fundamental concept in physics pervasive in various fields ranging from ferromagnetism and nematic phases in condensed matter to the Higgs mechanism in particle physics [1-3]. In condensed matter systems, the emergence of symmetry-broken phases is often associated with many-body interactions, with nematic order being a prime example. The nematic phase occurs when a quantum state breaks the rotational symmetry of the lattice, while maintaining its translational symmetry. Since these phases are found universally in correlated systems [4-9,18,24,25], understanding their underlying mechanism is important to unraveling the many-body phenomena of quantum materials. A critical factor that can drive the emergence of nematicity is the fermiology of the material. In the context of Landau's Fermi liquid theory [10-12], it is suggested that the inherent instability of electronic states can lead to the spontaneous breaking of point group symmetry and, consequently, the deformation of the Fermi surface, which is known as the Pomeranchuk instability [11]. In real materials, the circumstances are often complicated due to the coexistence of multiple degrees of freedom such as spin, orbitals and lattice, and experimental observations of Fermi surface deformations have been elusive [9,37-40].

Recently, Kagome materials have received tremendous attention due to its rich phase diagram, including superconductivity[16,17], unconventional charge density wave[13-15] and electronic nematicity[18]. Here we directly image an intra-unit-cell nematic order on the Kagome lattice via scanning tunneling microscopy (STM). This nematic order, which is tied to the Kagome lattice, breaks the rotational symmetry while preserving the translational symmetry, making it distinct from symmetry breaking orders found in the other Kagome materials [13-20,23,47]. Temperature-dependent studies reveal that the nematic order emerged below 70K on the Kagome layer. Notably, using spectroscopic maps we observe VHSs bands and their rotational symmetry breaking manners. Specifically, the van-Hove bands appear only around one M point of the Brillouin zone, while they are absent around other high-symmetry M points. Moreover, we resolve an elliptical band in close vicinity to the Fermi level, indicating the occurrence of the electronically mediated rotational symmetry breaking. Our results provide experimental signatures for VHS-driven nematicity in the Kagome lattice and demonstrate a complex interplay of VHS, nematicity, and multi-orbital physics.

$ScV_6Sn_6$ possesses a hexagonal structure [space group *P6/mmm* (No. 191)], composed of Kagome bilayers of Vanadium (V) atoms, and multiple layers of Tin (Sn) and Scandium (Sc) atoms, as illustrated in Fig. 1**a**. The compound has three different types of Sn sites: $Sn_1$, which forms a honeycomb lattice, $Sn_2$, which forms a triangular lattice, and $Sn_3$, which forms a honeycomb lattice with Sc atoms at the center [28]. To study the electronic structure of $ScV_6Sn_6$, we cleave the sample along the ab plane. After an extensive search, we have found two terminations with large areas and stable tunneling conditions. The topography for the two terminations, obtained by scanning tunneling microscopy (STM), is shown in Figs. 1**b** and 1**c**, acquired at 4.2K. Combining topographic and spectroscopic results, we identify termination A as the $Sn_1$ honeycomb layer and termination B as the V Kagome layer [see SI]. The topographic image on the $Sn_1$ layer (Fig. 1**b**) shows a honeycomb lattice with individual atoms clearly resolved. Strikingly, the topographic image on the Kagome layer exhibits a striped pattern breaking the six-fold rotational symmetry within the Kagome unit cell, while preserving the lattice translational symmetry (Fig. 1**c** inset) and we attribute it to an intra-unit-cell nematic order on the Kagome lattice. To confirm the existence of the nematic order, we conduct temperature dependence measurements. As shown in Fig. 1**d**, the striped pattern on Kagome layer persists up to 50K and disappears above 70K. The nematic order can also be identified from the Fourier-transform of the topographic images. Figure 1**e** shows the Fourier transform of topography on V Kagome and $Sn_1$ honeycomb layers respectively, which are conducted at 4.2K. Interestingly, they exhibit quite different characteristics. Firstly, on the Kagome layer, the intensity of Bragg peaks, as marked by green circles, exhibits anisotropic features, with the one along $q_1$ much stronger than those along $q_{2,3}$. This *k*-space anisotropy is a direct manifestation of the nematic stripe modulations in the real space topography data. In contrast, the Bragg peaks on the $Sn_1$ honeycomb layer are almost isotropic. Secondly, on the $Sn_1$ layer, we observe a set of charge density wave (CDW) peaks while they are absent on the Kagome layer. The CDW peaks correspond to a $\sqrt{3} \times \sqrt{3}$ superlattice modulation in the *ab* plane of the crystal, which is consistent with existing reports of CDW transition at $T_{CDW}$ = 92K in ScV$_6$Sn$_6$ [28-34]. It is worth noting that first-principles calculations show an imaginary phonon branch originating from Sn and Sc atoms [29]. Recently, there is also accumulating evidence showing that the CDW is related to the softening of phonon modes of Sn and Sc atoms [41-46]. Thus, our data support that the CDW is driven by a lattice instability limited to the Sn and Sc layers and less associated with the Kagome lattice.

Temperature dependent measurements further demonstrate that the nematic orders are less associated to the lattice reconstruction of the CDW phase. Figure 1**g** tracks the intensity of Bragg peaks on the Kagome layer at different temperatures, where the directions of linecuts are labeled in Fig.1**e**. Figure 1**f** further shows the Fourier transform image at 70K on both layers. On Kagome layer, the Bragg peaks show isotropic intensity at 70K, indicative of the destruction of nematic order, while the CDW still exists on honeycomb layer. Therefore, the critical temperature for the nematic order is below the CDW transition temperature. Since the two orders happen at different layers and different temperatures, we conclude that they are less associated to each other. On the Kagome lattice, the rotational symmetry breaking of Bragg peaks and absence of CDW peaks both point to an intra-unit-cell nematic order preserving the translational symmetry, which has not been previously observed in Kagome materials. Our observation that the nematic order only appears on the Kagome lattice implies an underlying relation between nematicity and lattice geometry, which we will elucidate in the following.

Now we focus on the electronic nature of the nematic order. From the topographic images, we observe that the nematic stripe patterns systematically emerge at low bias voltage. At 300 mV (Fig. 2**a**), it shows a honeycomb-like lattice without atomic resolution, akin to reported STM topographic images on Kagome layers of other 166 compounds [21,22]. At lower bias voltage, the stripes become prominent, as the schematic in Fig. 2**b**. STM topographic image is a convolution of lattice corrugation and electron density modulation [25]. If the stripe pattern comes from the lattice distortions, it should be captured by scanning at high bias voltage as well. Therefore, the prominence of nematic order at low energy suggests it is an electronic nematic order. We further conducted a thorough STS mapping to obtain information on the modulation of the electron density of states in both real and momentum space. Figure 2**c-d** displays the intensity of three distinct Bragg peaks as a function of energy, acquired from Fourier-transformed STS maps. Notably, on the V layer (Fig. 2**c**), one branch of Bragg peaks exhibits significant differences from the other two branches at most energy levels, indicating a breaking of rotational symmetry. However, the intensity of symmetry breaking on the Sn$_1$ layer becomes much weaker. Furthermore, the anisotropy of Bragg peaks exhibits an energy-dependent behavior, with the strongest intensity slightly above the Fermi level. The unusual energy dependence of the nematicity suggests that it is correlated with some degree of freedom at the same energy scale. Figure 2**e** shows the point dI/dV spectroscopy, which probes the local density of states. Interestingly, we observe a dI/dV peak at approximately $E_p$ =14 meV above the Fermi level. Figure 2**f** captures an intensity plot for a line of dI/dV spectra across the nematicity, demonstrating the robustness of the peak. By taking a one-dimensional Fourier transform of the dI/dV line, we find the strength of the modulation (at the nematic wavevector) to be strongest at the peak energy. The connection between the nematicity and electronic density of states raises a natural question – is the intra-unit-cell nematicity an electronic order? To answer this question, we first need to shed light on the origin of the peak in the density of states.

We now demonstrate that the peak in the density of states at $E_p$ corresponds to the saddle point of the VHS on the Kagome lattice. Quasiparticle interference (QPI) maps obtained from Fourier-transformed STS maps can be used to reveal the modulation of the local density of states and allow to determine the quasiparticle momentum dependence. Figure 3**a** displays a series of QPI maps at energies around $E_p$. At -8 meV, we observe a pair of arcs near the $\Gamma$ point. Each arc is mirror symmetric along the $\Gamma - M$ axis. At higher energies, the arcs move close to $\Gamma$, while remaining centered on the $\Gamma - M$ axis. At 14.5 meV (around the $E_p$), the arcs cross and move away from each other at higher energy along $\Gamma - K$ axis. The arc feature in QPI possibly stems from the scattering between two branches of van Hove bands [26], as shown schematically in Fig. 3**b**. To confirm this, we extract linecuts from QPI maps at various energies to obtain a QPI dispersion, shown in Fig. 3**c**. The dispersion exhibits a hole branch along the $\Gamma - M$ direction and an electron branch along $\Gamma - K$, consistent with the presence of a VHS. The two branches merge at around 15meV, in good agreement with the location of the conductance peak. The correspondence between QPI maps and conductance spectroscopy strongly supports the observation of a VHS.

Importantly, we observe that the QPI patterns break the rotational symmetry. Since the crystal structure of $ScV_6Sn_6$ exhibits six-fold rotational symmetry, the VHSs should appear at three inequivalent M points [29]. However, as shown in Fig. 3**a**, from the QPI maps we find that the features of van-Hove bands around the $\Gamma$ point break rotational symmetry: They preserve the mirror symmetry along the $\Gamma - M$ axis, while breaking the mirror symmetries along the $\Gamma - M'$ and $\Gamma - M''$ axis. The rotational symmetry breaking of QPI patterns can come from the anisotropic scattering potential, in which case the symmetry breaking axis is usually tied to the defect[48]. However, in our case, the interference patterns share the same symmetry axis, as shown in the real space dI/dV maps (Fig. 3**d** and Fig. S13). Therefore, the symmetry breaking of VHSs are more likely from sample's intrinsic property. The disappearance of VHSs at two of three inequivalent $M$ points indicates that the VHSs are closely related to the nematic order.

To further investigate the symmetry breaking of electronic states, we conduct QPI mapping over a large energy range. Notably, we observe a ring-shaped feature surrounding the center of Fourier-transformed STS maps as shown in Fig. 4**a-c**. The dispersion of the ring is extracted from cuts labeled in Fig. 4**a** and is displayed in Fig. 4**d**. The dispersive band persists from 300meV to the Fermi level, with significant band broadening observed around the VHS, indicating the effects of electronic interactions [27]. To identify the origin of QPI signals, we calculate the $ScV_6Sn_6$ band structure as shown in Fig. 4**e**. After considering the length and dispersion of scattering vectors, we attribute the elliptical QPI signals (Fig.4**a-c**) to a cone-like band around K point, which has dxz and dyz orbital characters. The constant energy cut at 100 mV (Fig. 4**f**) schematics the origin of scattering vector $V_2$ and $V_1$ drawn in Fig. 4**b** (see Fig.S17 for cuts at other energy). Remarkably, the QPI ring features an elliptical shape, breaking the six-fold rotational symmetry, as shown in Fig. 4**a**. In Fig. 4**g**, we plot the length of the QPI vectors $V_1$ and $V_2$ along the direction of $\Gamma - M$ and $\Gamma - M'$, marked by the blue and yellow arrows in Fig. 4**b**. It shows a persistent anisotropy throughout the entire energy window. To quantify the strength of symmetry breaking, we calculate the asymmetry ratio $\gamma$, defined as the ratio of the amplitudes of the vectors $V_2$ and $V_1$. In a six-fold rotationally symmetric system, one should expect no asymmetry, i.e., $\gamma = 1$. Plotting $\gamma$ as a function of the energy shows an increase trend at lower energy (Fig. 4**h**), indicating the band distortion becomes more dramatic toward the Fermi level.

At this stage, it is important to examine the possible extrinsic origin of the rotational symmetry breaking states. In STM experiments, anisotropic STM tips can produce asymmetric patterns since the measured results represent a convolution of the tip's LDOS and the sample's properties. Here, we observe the k-space distortion of electronic band, which cannot result from tip effects. The tip effect can be further ruled out by the nematic domains along different directions (see Fig. S15). Another possible origin is structural change. To test whether there are structural changes between 50K and 70K where we see the nematic transition, we performed single crystal x-ray diffraction measurements. The observed intensities agree well with the calculated intensities of the same structure at 50K and 70K (see Fig.S14 in SI) This indicates no structural transition between 50K and 70K. We note that while the electronic band is deformed in *k* space, we do not observe the distortion of Bragg peaks. To quantify the possible distortion of Bragg peaks, we extract the modulus of Bragg wavevectors $|Q_{1,2,3}|$ and plot the ratio of the wavevector $|Q_3|$ (along nematic direction) to others, as shown in Fig. 4**i**. Within our detection limit (1%), we do not observe any noticeable trends. In contrast, the distortion of electronic band is dramatic (exceeds 35%). Therefore, we conclude that within our detection limit (1%), the crystal lattice structure preserves the six-fold rotational symmetry, and that the deformation of Fermi surface is likely of electronic origin rather than lattice instability.

Building upon these observations, we aim to gain a deeper understanding of the origin of the nematic order on the Kagome layer. To this end, we disregard the influence of CDW ordering on nearby lattice sheets and instead focus on the electronic states without spin-orbit coupling. We conduct first-principles calculations to investigate the electronic structure of $ScV_6Sn_6$. As shown in Fig. 5**a**, the low-energy band structure of

ScV$_6$Sn$_6$ is dominated by V $d$-orbitals, forming a pair of VHSs at the $M$-point and a Dirac cone feature along the $M - K$ line. Orbital-resolved calculations suggest that they originate from d$_{z2}$ and d$_{xy/x2-y2}$ orbitals of V atoms. The pair of VHSs have different dispersions, and since we observe a hole branch along the $\Gamma - M$ direction in VHS-related QPI signals (Fig.3), it is most likely attributed to the VHS with d$_{xy/x2-y2}$ character. Furthermore, the Dirac crossing is protected by the opposite mirror eigenvalues of the two bands, resulting from their distinct orbital contribution: one is formed by V d$_{z2}$-orbitals, while the other is of V d$_{xy}$ character [29,46].

The electronic structure is reminiscent of the V-Kagome compound AV$_3$Sb$_5$, where a pair of VHS of sublattice-pure and sublattice-mixed type near the Fermi level induces an unconventional CDW order[15,20]. These VHSs have support on a single sublattice and pair of sublattices respectively, promoting long-ranged interactions [35-37]. However, ScV$_6$Sn$_6$ exhibits a different type of VHS, with the two branches of V-Kagome bands forming a pair of pure-type VHSs which both possess support only on a single sublattice. As a result, a CDW promoted by the sublattice-interference mechanism is absent. Instead, the observed nematic distortion of the Fermi surface may be due to an electronic instability, where the lowered symmetry of the distorted Fermi surface reduces the density of states at the Fermi level. To hybridize the bands at the Dirac point and shift the VHSs to higher energies, we consider a nematic order that breaks the mirror symmetries of the Kagome lattice. We model the instability by adding an on-site orbital hybridization term to two sublattices in a minimal model for the V-d$_{xy/x2-y2}$ and V-d$_{z2}$ orbitals, effectively capturing the instability (see Fig. 5**b**, blue) and reproducing the symmetry properties of the undistorted band structure (see Fig. 5**b**, red). Due to the pure sublattice character of the VHSs, this hybridization term modifies the states at two of the three inequivalent $M$-points, shifting the bands away from the Fermi level (see Fig. 5**b-c**, blue) and explaining the elliptical pocket observed in the QPI maps. Concomitantly, the VHS without support on the modified sublattices remains untouched, breaking the six-fold rotational symmetry of the underlying Kagome lattice and agreeing well with the experimentally observed nematicity on the V-layer, showing the disappearance of two of the three VHSs.

Electronic nematic order in correlated materials can arise through various mechanisms[7]. One most studied example is the nematic order observed in iron-based superconductors, where spin fluctuations and orbital orders have been considered essential[8,9]. Besides, novel mechanisms such as finite-energy nematic scenario have also been proposed for systems with low density of states near the Fermi surface, such as BaNiS$_2$[50]. In our data, the deformed Fermi surface and the nearby van-Hove singularity provide a key signature for inherently electronic instability. The non-observation of lattice distortions and structural transitions further constrains the origin of electronic nematicity and supports an intrinsically electronic mechanism. Theoretically, the VHSs near the Fermi level provide a large density of states, promoting the occurrence of electronic instability. In line with this, our model phenomenologically captures the symmetry lowering of the Fermi surface. The multi-orbital nature of VHSs ensures that multiple VHSs occur at similar energy and momentum without hybridizing at the single-particle level due to orbital orthogonality. By lowering the point group symmetry, hybridization between VHSs is enabled and therefore leads to the efficient removal of density of states from the Fermi level. Microscopically, interaction-driven hybridization of VHSs is known to be energetically favored at weak coupling when the two VHS have opposite concavity[49], consistent with the band structure of $ScV_6Sn_6$.

Finally, we discuss the nematicity observed in $ScV_6Sn_6$ and in the family of Kagome superconductors $AV_3Sb_5$[14]. In $CsV_3Sb_5$, electronic nematicity has been observed using nuclear magnetic resonance (NMR), elastoresistance measurement and STM[18]. Figure 5d represents part of the Fermi surface in $AV_3Sb_5$ schematically. There are three van-Hove singularities sitting at the inequivalent M points. Nesting vectors connecting the van-Hove singularities match with the 2x2 charge order, and the nematic order is observed at this finite **Q** order parameter, as shown in STM experiments[18]. In other words, the nematic order coexists with a 2x2 charge density wave, breaking the translational symmetry of the material. However, in $ScV_6Sn_6$, the $\sqrt{3} \times \sqrt{3}$ charge order does not fit the nesting scenario. Instead, we observed an intra-unit-cell order breaking the rotational symmetry while preserving the translational symmetry. The nematic order here carries no momentum and only breaks the point group symmetry, making it distinct from the nematicity seen in $AV_3Sb_5$ – akin to the scenario for chiral order proposed recently[49]. Besides, in $AV_3Sb_5$, multiple factors have been considered important for the electronic nematicity, such as local strain[51] and charge density wave[18]. In $ScV_6Sn_6$, our STM and X-ray data constrains the lattice distortion below 1%. The topographic images show flat surface without wrinkles, typically seen on strained sample (see Fig.S16), which limits the effects of local strain. Future experiments such as strain-free transport measurements[51], nematic susceptibility and NMR[18] would be invaluable to probe the phase diagram of nematic order and unveil its origin.

**Acknowledgement:**

M.Z.H. acknowledges support from the US Department of Energy, Office of Science, National Quantum Information Science Research Centers, Quantum Science Center and Princeton University. Theoretical and STM works at Princeton University were supported by the Gordon and Betty Moore Foundation (GBMF9461; GBMF4547; M.Z.H.). Work at Nanyang Technological University was supported by the National Research Foundation, Singapore under its Fellowship Award (NRF-NRFF13-2021-0010) and the Nanyang Technological University start-up grant (NTUSUG). M.M.D. acknowledges support from a Forschungskredit of the University of Zurich (Grant No. FK-22-085). T. N., M.M.D., and S.B.Z were supported by the European Research Council (ERC) under the European Union's Horizon 2020 research and innovation program (ERC-StG-Neupert-757867-PARATOP). S.B.Z was also supported by the UZH Postdoc Grant. R.T. was supported by the Deutsche Forschung gemeinschaft (DFG, German Research Foundation) through Project-ID 258499086- SFB 1170 and the Wurzburg-Dresden Cluster of Excellence on Complexity and Topology in Quantum Matter – ct.qmat Project-ID 390858490-EXC 2147. Y.G. acknowledges the Double First-Class Initiative Fund of ShanghaiTech University. W. X. acknowledges the research fund from the State Key Laboratory of Surface Physics and Department of Physics, Fudan University (Grant No. KF2022_13). Y.Y.P. is grateful for financial support from the National Natural Science Foundation of China (Grants No. 12374143).

**Competing interests:** The authors declare no competing interests.

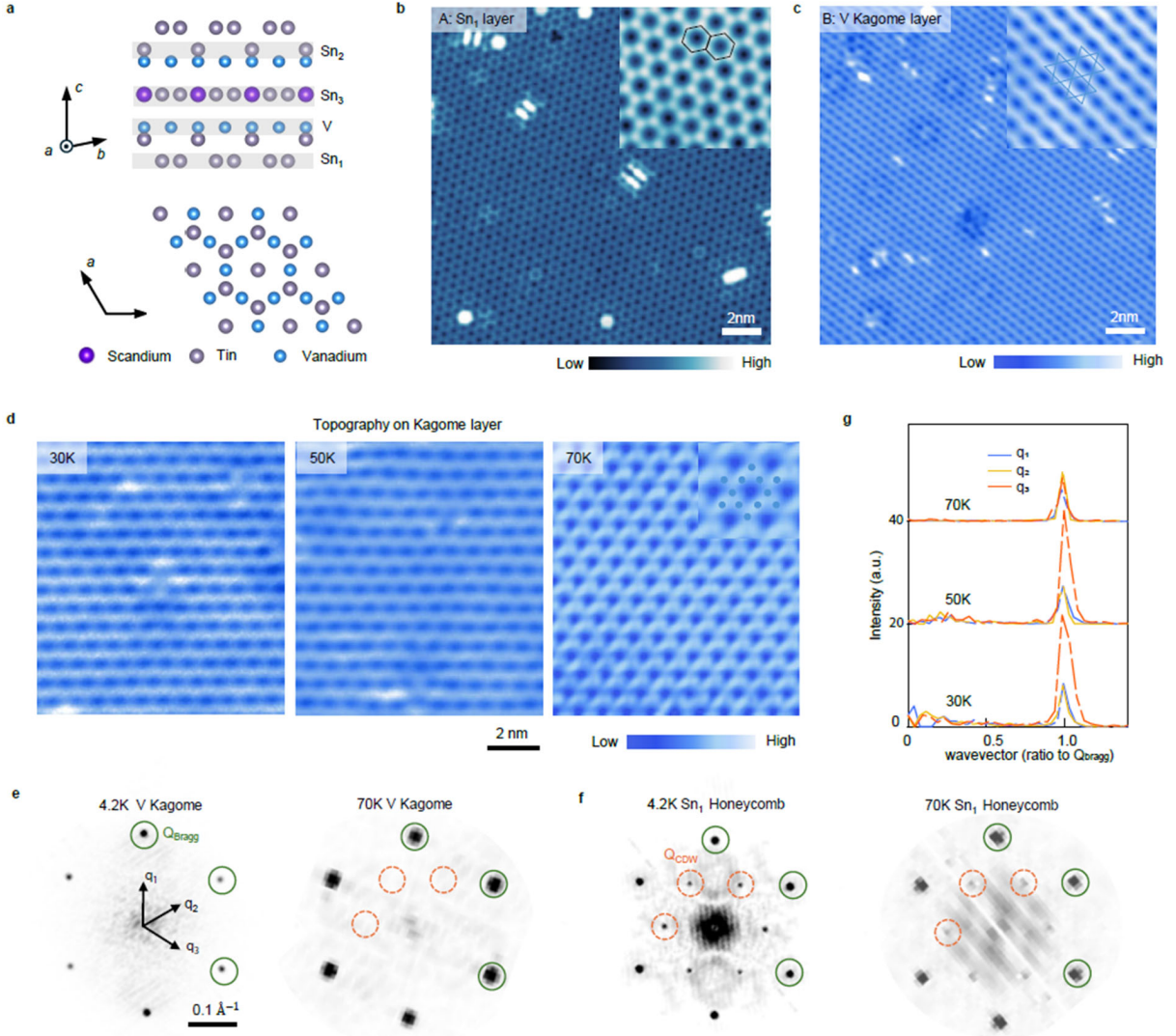

**Figure 1. Intra-unit-cell nematic order and charge density wave in ScV$_6$Sn$_6$. a,** Crystal structure of ScV$_6$Sn$_6$. Multiple possible terminations are highlighted. The lower panel shows the top view of the lattice. **b,** STM topographic image of the Sn$_1$ layer. The inset shows the atomically resolved Sn$_1$ honeycomb lattice. **c,** Topographic image of the V Kagome layer. The inset shows the stripe pattern and the corresponding Kagome lattice structure. Tunneling junction setup for B and C: V = 100mV, I = 50pA (1nA for the inset). **d,** Topography of the Kagome layer at different temperature. An intra-unit-cell nematic order emerges below 70K. **e,** Fourier-transform of the topographic image of the Kagome (left) and honeycomb (right) plane at 4.2K. Bragg peaks are labeled by green circles. $\sqrt{3} \times \sqrt{3}$ CDW peaks are highlighted by orange circles. **f,** Fourier-transform of the topographic image of the Kagome (left) and honeycomb (right) plane at 70K. **g,** Temperature dependent Fourier-transform linecuts on the Kagome layer. Directions are along $q_1$, $q_2$, $q_3$ crossing the Bragg peaks which are denoted in panel E.

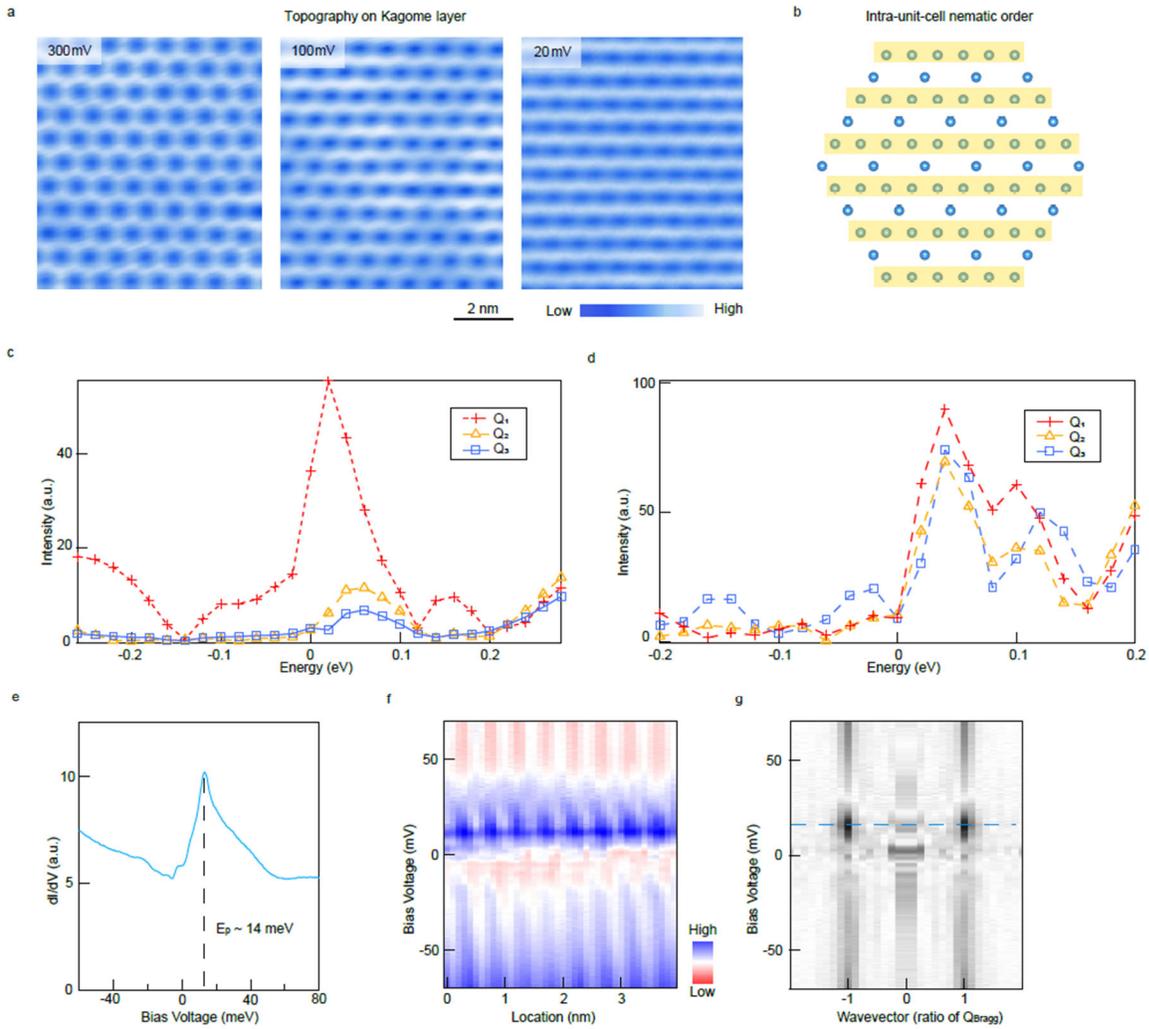

**Figure 2. Evidence of electronic nematicity. a,** Topography of the Kagome plane under different bias voltage at 4.2K. The intra-unit-cell nematic order gradually emerges at 100meV and 20meV. **b,** Schematics of the intra-unit-cell nematic order on the Kagome lattice. **c,** Energy-resolved intensity of Bragg peaks $Q_{1,2,3}$ on honeycomb (d) and Kagome (c) terminations, extracted from Fourier-transformed dI/dV spectroscopic maps. **e,** Averaged dI/dV spectroscopy on the Kagome plane. There is a peak in dI/dV that is located at around 14meV, suggesting a pronounced density of states. **f,** Intensity plot of a dI/dV spectroscopy linecut across the direction of the nematicity. Tunneling junction setup: V = 80mV, I = 1nA. **g,** One-dimensional Fourier transform of the line spectroscopy. Vertical lines at wavevector = ±1 correspond to the modulation at the nematicity vector. The blue dashed line marks the energetic position of the dI/dV peak, where the modulation is found to be the strongest. Wavevectors are shown in the unit of vectors for Bragg peaks.

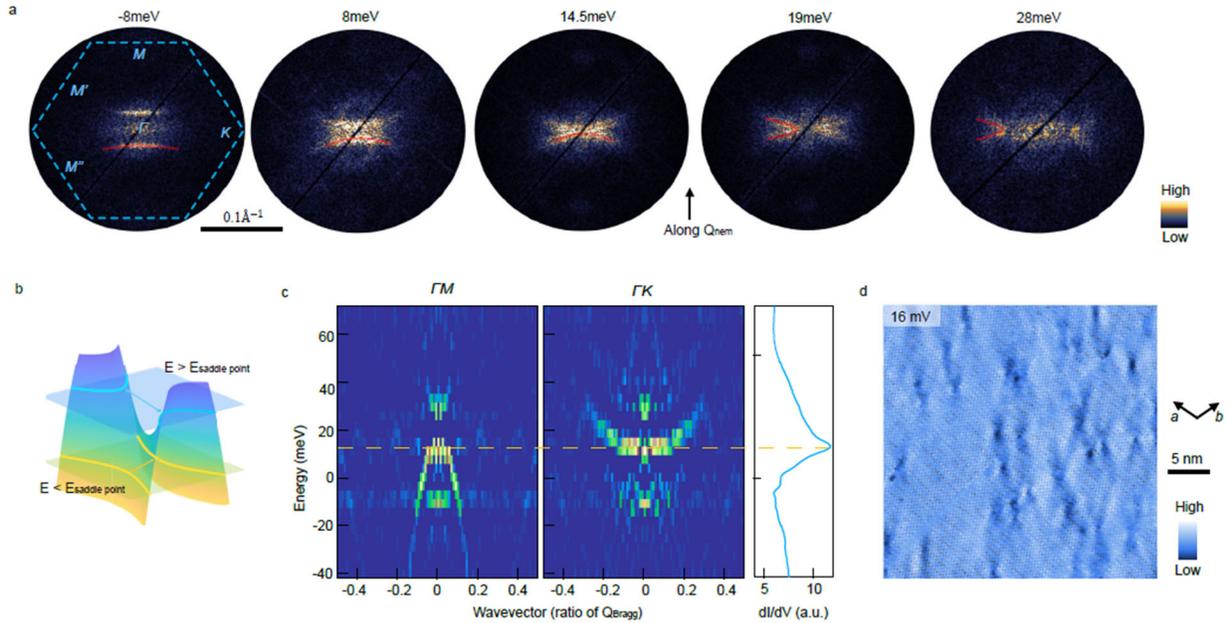

**Figure 3. Observation of van-Hove singularities and their annihilation. a,** Quasiparticle interference (QPI) maps at the corresponding energy layers (relative to the Fermi level). The blue dashed line marks the first Brillouin zone centered on Γ. The red curves guide the scattering features from the VHS bands. **b,** Schematic of electronic bands near the VHS. Blue and yellow curves are the constant energy contour above and below the saddle point energy. The arrows illustrate the quasiparticle scattering between VHS bands. **c,** QPI dispersion along the directions of Γ − M (left panel) and Γ − K (right panel), where the momentum points are labeled in **a**. The yellow dashed line marks the energetic position of dI/dV peak (shown in Fig. 2**e**), which matches the position of the saddle point. Wavevectors are in the unit of vectors for Bragg peaks. **d,** Real space dI/dV map at 16 mV.

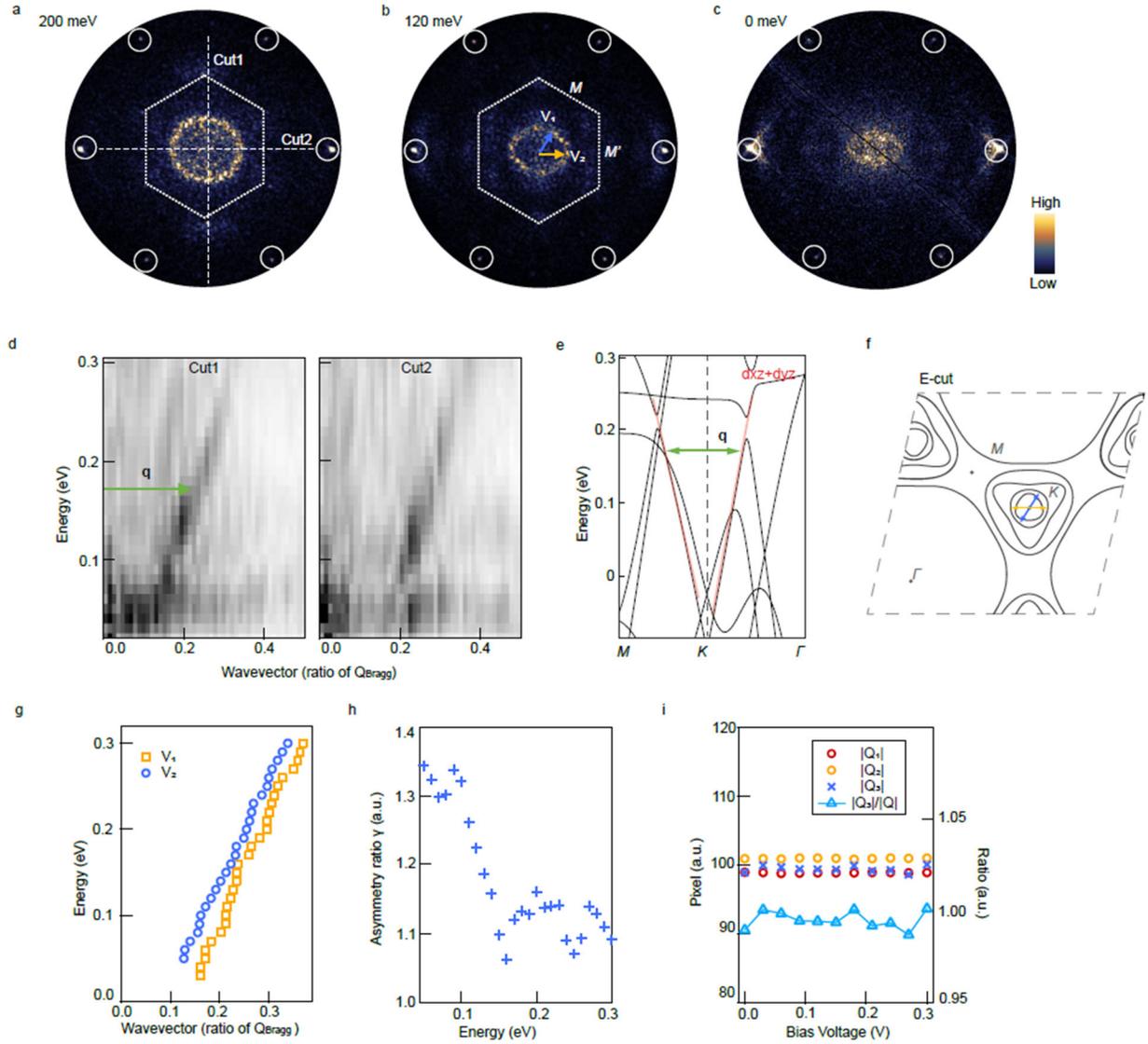

**Figure 4. Deformation of the low-energy electronic states. a-c,** QPI data at 200 meV, 120 meV and 0 meV. An elliptical ring appears at the center, representing the scattering of electronic states. Bragg peaks are marked by white circles. BZ is labeled. **d,** Linecuts extracted from QPI, showing the quasi-particle dispersion. Directions of cut1 and cut2 are marked in a. **e,** DFT calculated band structure. Bands related to QPI dispersions in **d** are highlighted. q denotes the origin of QPI scattering vector labeled in **d**. **f,** Calculated constant energy cut at 100 meV (kz=0). The Origins of scattering vectors V1 and V2 (in **b**) arere labeled. **g,** Dispersion of QPI ring along $\Gamma - M$ and $\Gamma - M'$. V1 and V2 are the momentum vectors of QPI ring along directions illustrated in **b**. The QPI ring has a finite bandwidth. At each energy, the magnitude of V1 and V2 are extracted by taking the median value of the bandwidth. **h,** Asymmetry ratio of the QPI ring γ, defined as the ratio of the magnitude of V2 to V1, plotted as a function of energy. The distortion of electronic bands becomes more dramatic at lower energy. **i,** Modulus of three Bragg wavevectors. $Q_3$ is along nematic direction. Ratio of $|Q_3|$ to $|Q|$ (average) fluctuates within 1% without noticeable trend.

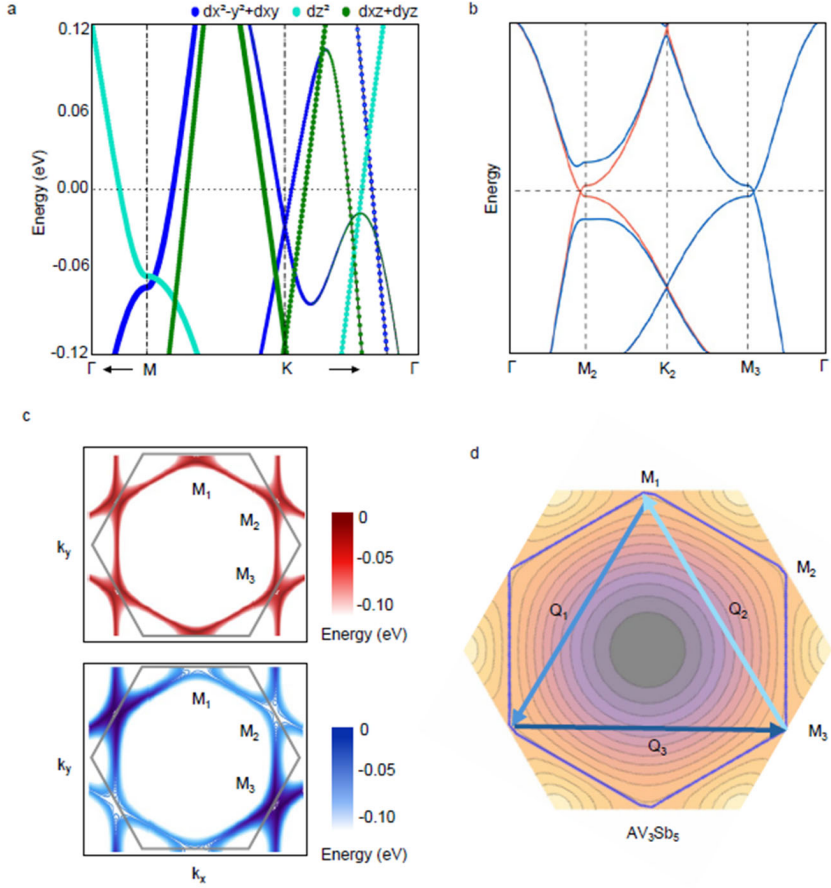

**Figure 5. Impact of nematic instability on the electronic structure of $ScV_6Sn_6$. a,** Orbital resolved band structure of $ScV_6Sn_6$. **b,** Low-energy effective band structure of V $d_{xy}/d_{x2-y2}$ and V $d_{z2}$ orbitals for the normal (red) and nematic (blue) state. The normal state shows a pair of VHSs above and below the Fermi level at each M-point, forming a mirror-symmetry protected Dirac crossing along the M-K line. The nematic phase results from electronic instability, breaking all mirror symmetries and gapping the Fermi surface around two of the three inequivalent M-points. **c,** Constant energy cuts at and below the Fermi surface, highlighting the evolution from a six-fold rotationally symmetric Fermi surface in the normal state (red, top) to the nematic instability (blue, bottom), breaking the rotational symmetry of the Kagome lattice and retaining the original VHSs only at a single M-point. **d,** Energy contour plot and schematic partial Fermi surface (blue) of $AV_3Sb_5$. Van Hove singularities sit at three inequivalent M-points (M1, M2, M3). The 2x2 charge order peaks match the nesting vectors ($Q_{1,2,3}$) connecting the van Hove singularities match; anisotropic features are observed at finite-Q vectors, unlike the electronic instability in $ScV_6Sn_6$ which occurs at zero wavevector.

# Methods

**Crystal growth**

The ScV$_6$Sn$_6$ crystals were grown by using the self-flux method. Sc (99.9%) blocks, V (99.9%) powder and Sn (99.999%) granules were mixed in a molar ratio of 1: 6: 40 and put into an alumina crucible. The crucible was sealed into a quartz tube in vacuum and was subsequently heated in a furnace up to 1150 °C in 15 hrs. After reaction at this temperature for 10 hrs, the assembly was cooled down to 750 °C within 150 hrs. The excess Sn was quickly removed at this temperature in a centrifuge, and black crystals with shining surface and a typical size of 1.5×1.5×0.5 mm$^3$ were left.

**STM experiments:**

ScV$_6$Sn$_6$ single crystals were cleaved at 78K under ultrahigh vacuum (UHV) conditions ($< 5 \times 10^{-10}$ mbar), and then immediately inserted into the microscope head, already at the $^4$He base temperature. The cleaved sample shows a [001] surface with atomic flatness. In this study, we have tried more than 20 samples from two batches. Among 7 successfully cleaved samples, we have found clean Kagome termination with atomic resolution in 3 samples (two from batch A and one from batch B). In these 3 samples, we have examined over 30 surfaces, and the nematic patterns appear in all the Kagome terminations. Commercial Platinum-Iridium (Pt-Ir) tips were annealed in UHV camber and then characterized with a reference sample. Tunneling conductance maps were obtained using standard lock-in amplifier techniques with a lock-in frequency of 974 Hz and tunneling junction set-ups as indicated in the corresponding figure captions. QPI dispersions in Fig.3 **c** are extracted using standard curvature methods for better visualization[38]. Raw data is provided in Supplementary figure 9. QPI maps in Fig.3 **a** are taken on 80nm × 80nm region with modulation voltage of 3 mV. QPI maps in Fig.3 **a** are taken on 50nm × 50nm region with modulation voltage of 12 mV. Before the Fourier transform, we subtracted a linear fit from the original real space image without harming the QPI signals. QPI data are presented without any symmetrization.

**X-ray diffraction**

Single crystal X-ray diffraction measurements were performed by using the custom-designed X-ray instrument. It is equipped with a Xenocs Genix3D Mo Kα (17.48 keV) x-ray source, which provides 2.5 × 10$^7$ photons/sec in a beam spot size of 150 μm at the sample position. The measured samples were mounted on a Huber 4-circle diffractometer and cooled by a closed-cycle cryostat with a beryllium dome. Diffraction signals are collected by a highly sensitive single-photon counting PILATUS3 R 1M solid state area detector with 981 × 1043 pixels. Each pixel size is 172 μm × 172 μm.

**First-principle calculations**

Our ab initio calculations are performed in the framework of density-functional theory within the Perdew-Burke-Ernzerhof exchange-correlation functional[52], as implemented in Vienna ab-initio Simulation Package (VASP)[53]. The projector augmented wave method [54] has been adopted with 3d14s2, 3d34s2 and 5s25p2 treated as valence electrons for Sc, V and Sn atoms, respectively. The cutoff energy of 400 eV and a dense k-point sampling of the Brillouin zone with a KSPACING parameter of 0.2 are used to ensure the enthalpy converged within 1 meV/atom.

**Data and materials availability:**

All data needed to evaluate the conclusions in the paper are present in the paper and the Supplementary Materials. Additional data are available from the corresponding authors upon reasonable request.